\documentclass{iau}
\usepackage{graphicx}
\usepackage{natbib}
\def\EBV{\mbox{E$_{\rm B-V}$}}

\def\AV{\mbox{A$_{\rm V}$}}

\def\HH{\mbox{H$_2$}}

\def\nH2{{\rm n}({\rm H}_2)}
\def\NH2{{\rm N}({\rm H}_2)}
\def\pccc{~{\rm cm}^{-3}} 
\def\pcc {~{\rm cm}^{-2}}

\def\Tstar#1 {\mbox{${\rm T}_{\rm #1}^*$}}
\def\Tsub#1 {\mbox{$T_{\rm #1}$}}

\def\TK  {\Tsub K }

\def\Texc {\Tsub exc }

\def\Tcmb{\Tsub CMB }

 \def\arcmin{\mbox{$^{\prime}$}}

\def\degr{$^{\rm o}$}
\def\p{\mbox{$^+$}}
\def\m{\mbox{$^-$}}

\def\hcop{\mbox{{HCO\p}}} 
\def\hocp{\mbox{{HOC\p}}} 
\def\chp{\mbox{CH\p}}
 
\def\cch{\mbox{C$_2$H}}
\def\cfh{\mbox{C$_4$H}}  
\def\hhco{\mbox{H$_2$CO}}
\def\h13cop{\mbox{{H$^{13}$CO\p}}}

\def\c3h2{\mbox{C$_3$H$_2$}}
 
 \def\R0{R$_0$}

\def\ddeg{{}^\circ\kern-.1em}

\def\kms{\mbox{km\,s$^{-1}$}}

\def\bll{BL Lac}

\def\E#1 {$10^{#1}$}
\def\E#1 {E{#1}}
\def\P#1,{$\nH2\TK~=~#1\times~10^4\pccc$~K}
\def\ec#1,#2,#3,{#1\,(#2)\E{#3}}

\def\methanol{\mbox{CH$_3$OH}}
\def\H3{\mbox{H$_3$}}

\def\ammon{\mbox{N\H3}}

\def\RH2{\mbox {R$_{\HH}$}}

\title[IAU 297~~Microwave probes of DIB carrier abundances] 
{What microwave astronomical spectroscopy \\ can tell you about the carriers of the DIBS}

\author[Harvey Liszt, Robert Lucas, Jerome Pety \& Maryvonne Gerin]  
{Harvey Liszt$^1$, Robert Lucas$^2$, Jerome Pety$^{3,4}$
 \& Maryvonne Gerin$^{5,4}$}

\affiliation{
$^1$Notional Radio Astronomy Observatory, 
520 Edgemont Road, Charlottesville, VA USA 22903 \\[\affilskip]
$^2$Institut de Planétologie et d.Astrophysique de Grenoble (UMR 5274)
BP 53 F-38041 Grenoble Cedex 9, France \\[\affilskip]
$^3$Institut de Radioastronomie Millim\'etrique, 300 Rue de la Piscine \\
F-38406 Saint Martin d'H\`eres, France   \\[\affilskip]
$^4$ Obs. de Paris, 61 av. de l'Observatoire, 75014, Paris, France \\[\affilskip]
$^5$LERMA/LRA, Ecole Normale Supérieure, 24 rue Lhomond, 75005 Paris, France}

\pubyear{2013}
\volume{297}  
\pagerange{119--126}
\jname{The Diffuse Interstellar Bands}
\editors{H. Linnartz  \& Jan Cami , eds.}

\begin{document}

\maketitle

\begin{abstract}

Astronomical microwave spectroscopy has greatly broadened the inventory
of identifiable chemical species in diffuse molecular gas and is an increasingly 
effective way to measure the abundances of polar molecules that may be
candidate diffuse interstellar band carriers.  
Here we review some recent developments that hold new promise for 
chemical abundance determinations.  We summarize and categorize the molecular 
inventory that has accrued in the past twenty years from  microwave observations 
of diffuse clouds and we present summary tables of the molecular abundances 
within various chemical families in both dark and diffuse molecular gas.
\end{abstract}

\firstsection 

\section{Introduction}

It is a paradox that the uv/optical spectrum is so rich in atomic lines and
diffuse interstellar bands (DIBS) and so empty of the identifiable 
molecular spectra that might finger the DIB carriers. CH, \chp\ and CN were 
identified in optical spectra when study of the interstellar medium was 
relatively new and it was expected that the carriers of the DIBS would 
soon also be identified.  But the known molecular inventory grew only 
very slowly in optical/uv spectra with the detection of \HH, CO, C$_2$, C$_3$ 
and NH \citep{SnoMcC06}.  The optical spectrum alone provides little guidance
and the DIB carriers remain unidentified.

By contrast, the microwave spectrum proved to be very fertile ground for
astrochemistry, with the detection of more than 150 interstellar 
species since 1963 when OH was discovered.  These are mostly studied in 
emission toward ``dark,'' ``giant,'' ``molecular,'' and  ``infrared dark'' 
clouds having visual extinctions \AV\ $>5$ mag. but substantial progress 
has also been made in detecting polyatomic molecules in regions with 
\AV\ $<$ 1 mag, in fact almost at the point where \HH\ itself ``turns on'' 
at reddening \EBV $\approx$ 0.08 mag or \AV\ $\approx$ 0.25 mag 
\citep{SavDra+77}.  Doing so only requires observing in absorption at radio 
wavelengths, as we discuss here.

Here we make the case for the usefulness of radioastronomical 
microwave spectroscopy to the quest for the carriers of the DIBS.  This 
rests on establishing several elements.  The spectroscopy must be 
capable of observing relevant species and setting significant limits on
their abundances.  Morever, it must be capable of observing those species 
in interstellar environments that are relevant to the DIBS.  The latter point
is perhaps the hardest to establish because stars are not suitable background
sources at radio wavelengths and the same sightlines are not observed in
both optical and microwave absorption (a possible future exception is discussed 
below). Largely on this account the optical/uv observing community has been very slow 
to acknowledge that diffuse clouds are actually observable in the microwave 
domain.  However, the recent microwave determination \citep{LisSon+12} of 
the abundance of the DIB carrier candidate {\it l}-\c3h2\ 
\citep{MaiWal+11} seems to have had some impact and may be seen as an 
initial step in cross-fertilization between the optical and microwave
domains.

The organization of this work is as follows.  Section 2 provides an 
exceedingly brief overview of the sensitivity to molecular abundance 
that is available in the microwave domain and an equally brief description
ot the kinds of transitions and molecular species that are accessible.  
It includes some specific examples of recent detections of new species in 
diffuse gas as a way of illustrating 
instrumental capabilities.  The systematics of the chemistry are
shown in Section 3 and the overall microwave molecular inventory is presented
in Section 4 along with summary tables of relative abundances in dark
and diffuse gas.  Section 5 provide a practical summary of the overall
relevance of this work to the quest for the identities of the DIB carriers.

\begin{figure}[a]
 \vspace*{2.0 cm}
\begin{center}
 \includegraphics[width=4in]{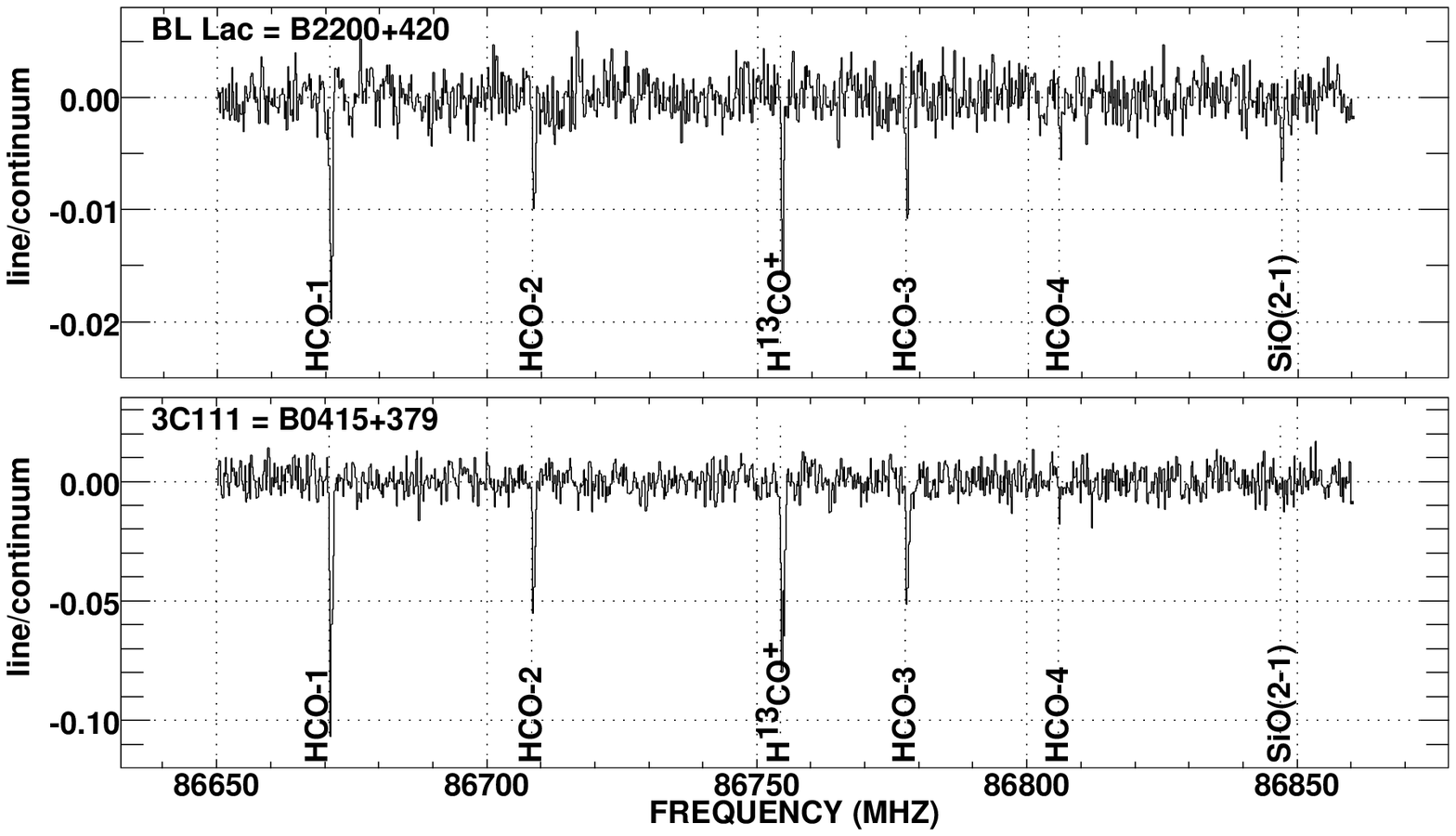} 
 \caption{MM-wave galactic absorption spectra toward two blazars, excerpted from 
  a 85-115 GHz spectral sweep using the EMIR receiver at the IRAM 30m telescope.  
  The spectral resolution is 195 kHz.}
   \label{fig1}
\end{center}
\end{figure}

\begin{figure}[b]
\begin{center}
 \includegraphics[width=4in]{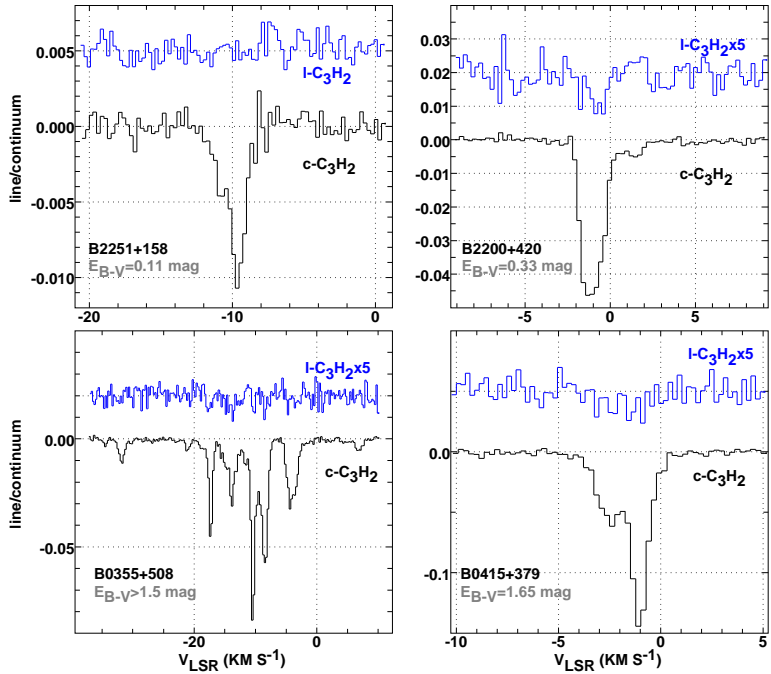} 
 \caption{19 GHz VLA absorption spectra of linear-para and ortho-cyclic \c3h2\ toward 
   four blazars, from \cite{LisSon+12}}
   \label{fig2}
\end{center}
\end{figure}

\begin{figure}[c]
\vspace*{2.0 cm}
\begin{center}
 \includegraphics[width=4.5in]{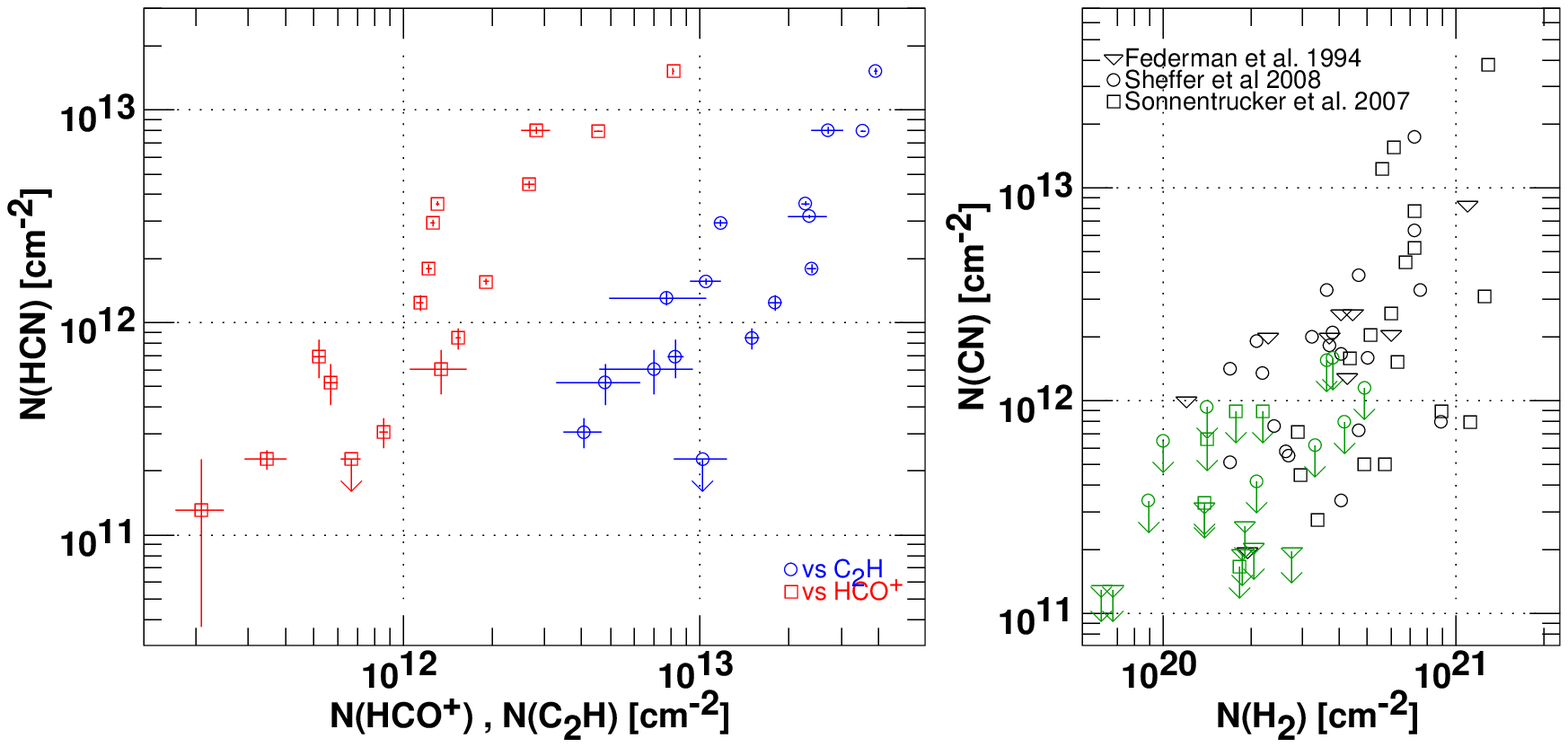} 
 \caption{Left: HCN column density variations determined at radio frequencies with respect to
   \hcop\ and \cch\ \citep{LisLuc01} . Right: Variation of N(CN) with N(\HH) from optical/uv spectra
    \citep{FedStr+94,SheRog+08,SonWel+07}.}
   \label{fig3}
\end{center}
\end{figure}

\begin{figure}[d]
\begin{center}
 \includegraphics[width=4.3in]{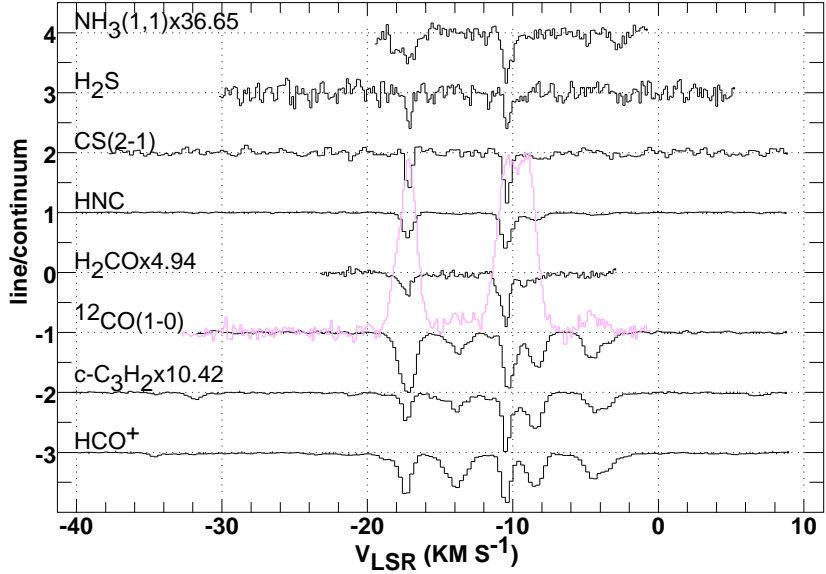} 
 \caption{Absorption spectra toward B0355+508 (see also Figure 2), 
  vertically offset and scaled as noted to fill more of the display. Shown as a faint
   cyan overlay just above the CO absorption spectrum is the CO J=1-0 emission profile 
  seen in this direction at 1\arcmin\ spatial resolution; the peak brightness is 3 K.}
   \label{fig4}
\end{center}
\end{figure}

\begin{figure}[e]
 \vspace*{2.0 cm}
\begin{center}
 \includegraphics[width=5in]{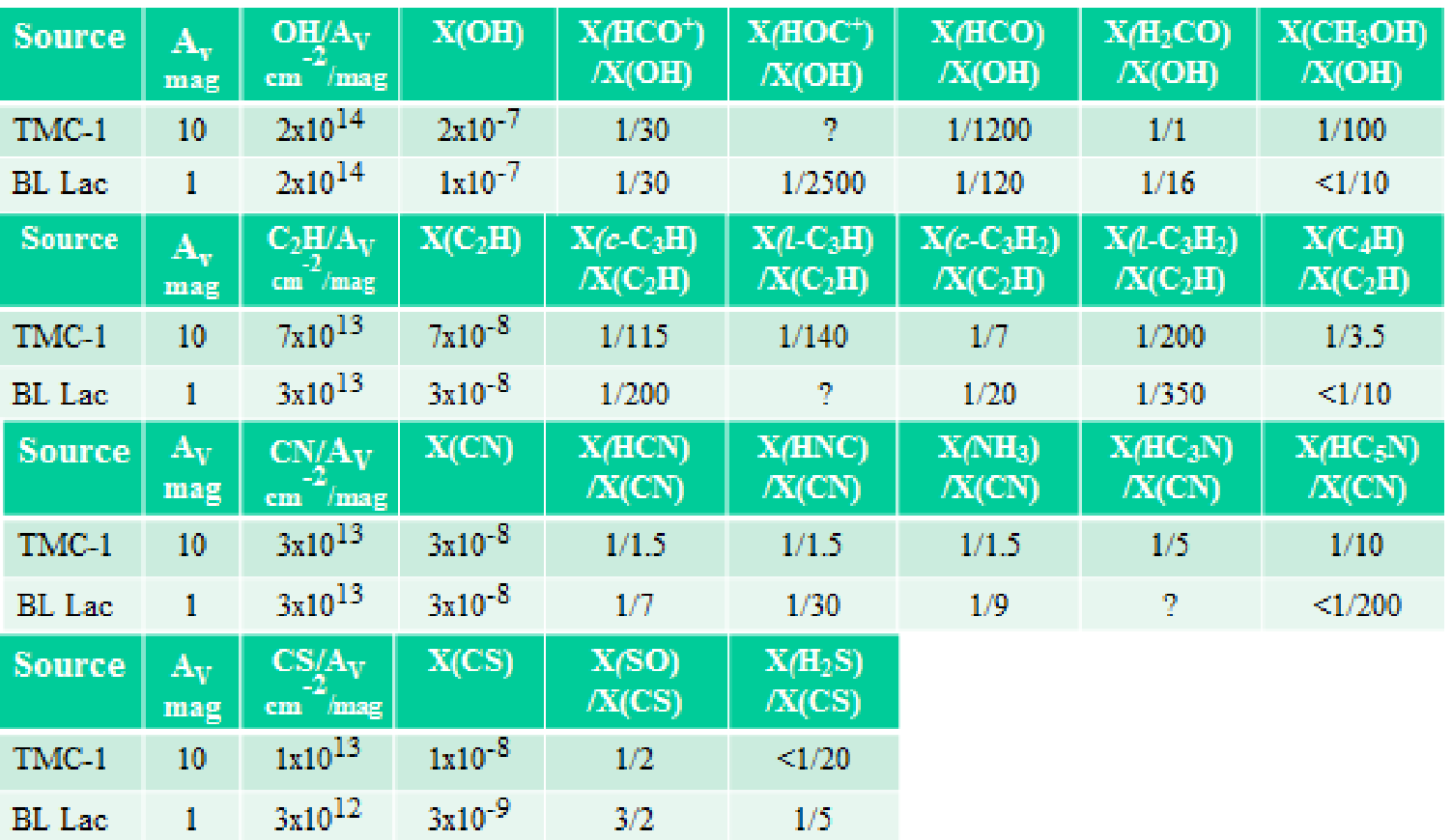} 
 \caption{Abundances toward TMC-1 \citep{OhiIrv+92,TurHer+00,SmiHer+04} and \bll.  
Except for CH, abundances relative to \HH\ toward \bll\ are relative to \hcop\ with
X(\hcop) = $3\times 10^{-9}$.}
   \label{fig5}
\end{center}
\end{figure}

\section{Considerations of observability and sensitivity}

\subsection{Sensitivity to column density}

There are two essential aspects of calculating the strength of an interstellar 
microwave absorption line.  First, as with optical spectroscopy, are the
particulars of the spectroscopic data for that particular transition, and for
 known interstellar species these can generally be found in radioastronomical 
spectroscopic databases {\footnote {\tt www.splatalogue.net, www.astro.uni-koeln.de/cdms, 
www.spec.jpl.nasa.gov}}. Second is the 
issue of rotational and fine-structure excitation within the ground vibrational 
state molecular energy level structure  and this is somewhat different
at radio wavelengths; substantial population exists in equilibrium in both the 
upper and lower levels of the transition and the molecular population is spread 
over more energy levels (and perhaps many more) than are actually observed.

For a transition at frequency $\nu$, between  lower and upper levels j and k 
having statistical weights g$_j$ and g$_k$ and an Einstein A-coefficient 
A$_{kj}$, the column density of molecules N$_j$ in the lower level j and 
the integrated optical depth (in cgs velocity units) are related as 

$$ \int \tau dv  = (c^3/(8 \pi\nu^3)) (g_k/g_j) A_{kj} (1-exp(-h\nu/k\Texc)) N_j $$

In this expression \Texc\ is the excitation temperature of the transition,
N$_k$/N$_j$ = (g$_k$/g$_j exp(-h\nu/k\Texc$).  Because $h\nu/k\Tcmb$ is not 
large ($h\nu/k\Tcmb = 1$ for $\nu = 57$ GHz) the correction 
for stimulated emission is usually important, appreciably lowering the 
optical depth and overall sensitivity.  Beyond this the molecular population 
is spread over various energy levels even when there is no appreciable 
excitation above the cosmic microwave background.  To derive the total 
column density N from a single measurement, the partition function 
(sum of population over all levels) must be calculated to evaluate the 
ratio Q=N/N$_l$.  The observability to a particular species at a given
radiotelescope can be optimized by choosing a transition that minimizes 
this ratio.  

In diffuse gas the ambient density is 
relatively small and the distribution of population over the rotational
energy ladder is well-characterized by taking \Texc = \Tcmb; this is 
verified by observing emission from species whose optical depths are known 
in absorption \citep{LucLis96,Lis12}.

\subsection{Observability of various species}

Heavier diatomic and smaller linear molecules (ie not OH, \ammon, \hhco) are 
generally observed in pure rotational transitions at mm-wave frequencies and 
have little population in rotational levels above J=2.  For a commonly 
observed linear species like \hcop\ with a comparatively large permanent
dipole moment 3.89 Debye and a J=1-0 transition at 89.188 GHz, an integrated 
optical depth of 1 \kms\ corresponds to N(\hcop) $= 1.12 \times 10^{12} \pcc$
when \Texc = \Tcmb.  Column densities N(\hcop) $ 10^{10} \pcc$ are 
easily accessible corresponding to molecular hydrogen column densities as 
small as N(\HH) $= 3 \times 10^{18} \pcc$.  As seen in Figure 1, the 
$^{13}$C-substituted variant of \hcop\ is easily detected toward \bll\
at \AV\ = 1 mag.

Heavier species have smaller rotational constants and transitions that lie
lower in frequency, with population spread over more levels. In terms of the 
partition function, \cfh\ is optimally observed in the N=2-1 transitions near 
19.015 GHz, and an integrated optical depth of 1 \kms\ corresponds only to 
N(\cfh) $= 7-10\times 10^{14} \pcc$ if the permanent dipole moment 
(which is somewhat uncertain) is as small as 0.8 Debye.  By contrast, the \cfh\m\ anion
with transitions at 18.6 GHz has a dipole moment of 6.2 Debye and a column density
sensitivity nearly two orders of magnitude better \citep{LisSon+12}.  

Transitions are also observable that are not purely rotational. Four hyperfine
transitions of OH are observed between 1612 and 1712 MHz and an
integrated optical depth of 1 \kms\ in the strongest of these at 1667 MHz
corresponds to N(OH) $\approx 3\times 10^{14}\pcc$; however OH is observed with 
very large instruments and column densities well below N(OH)$ =10^{13}\pcc$ are easily 
accessible \citep{LisLuc96}.  As an asymmetric rotor, \hhco\ is observed in rotational 
transitions at mm-wavelengths and in K-doubling transitions at 4.830 and 14.480 GHz
\citep{LisLuc+06}. \ammon\ is seen at 23.7 GHz because of inversion doubling of its 
ground-state pyramidal structure ({\it ibid}).  The pure rotational transitions of 
OH and \ammon\ are observable in diffuse gas in the sub-mm domain (above 300 GHz).
Column densities for these three species are shown in Figure 5.

Each species therefore must be taken on its own terms in terms of both observability
and sensitivity and no one radiotelescope is capable of observing the full panoply
of interstellar species.  However, nearly the entire microwave spectrum from 300 MHz to 
300 GHz, temporarily excepting 70 - 84 GHz (for instrumental reasons) and some opaque 
atomospheric windows, is observable with large modern instruments.

\subsection{Examples} 

Microwave radioastronomical spectra have always had relatively good velocity 
resolution but only recently has it become possible to observe with such high 
resolution over large fractional bandwidths.
Figure 1 shows the recent initial detection of galactic HCO in absorption at 
86.7 GHz against the blazar \bll\ at l,b = 92.6\degr,-10.44\degr\ with \EBV = 
0.33 mag and toward 3C111 at l,b = 161.7\degr,-8.8\degr\ with \EBV = 
1.65 mag; all reddenings quoted here are from \cite{SchFin+98}.
The absorbing gas seen in the 4 hyperfine components of the $1_{01}-0_{00}$ 
transition of HCO and in the H$^{13}$CO\p\ J=1-0 and SiO J=2-1 transitions is at v = 
-1 \kms\ in both directions in the LSRK velocity rest frame and lies 
within a few hundred pc of the Sun at these moderate galactic latitudes.  

These spectra were made by beam-switching against adjacent sky at the 
IRAM 30m telescope using the EMIR receiver that simultaneously observes over 
two 8 GHz-wide sidebands with 195 kHz spectral resolution, corresponding to 0.6 \kms\ 
at 100 GHz.  \bll\ was flaring at 15 Jy and 3C111 was anomalously weak
at 3 Jy during these observations, which represent a very small portion of 
the 32 GHz-wide spectrum constructed with several LO settings during some 25 
hours observing toward each source.  Toward \bll\ we derive  
N(HCO) $ = 8.5 \times 10^{11}\pcc$,  N(H$^{13}$CO\p) $ = 4.4 \times 10^{10}\pcc$ 
and N(SiO) = $ = 4.8 \times 10^{10}\pcc$.  The column densities of HCO and \hcop\ 
per unit reddening are very nearly the same in both directions but notice that SiO 
is weaker along the translucent sightline to 3C111.  SiO was previously observed
along most of the sightlines discussed here \cite{LucLis00} but the detection toward
\bll\ is new.

The observing bandwidths for high-resolution spectroscopy at the VLA are improving 
rapidly as a new spectrometer is commissioned.  Figure 2 from \cite{LisSon+12} 
shows absorption from the ortho-cyclic and para-linear versions of \c3h2\ made in 
4 hours (total) observing using the VLA at 18.5 GHz toward four blazars including 
those shown in Figure 1, but also toward B2251+158=3C454.3 at 
l,b = 86.1\degr,-38.2\degr\ with \EBV\ = 0.11 mag.  \c3h2\ is ubiquitous in 
diffuse molecular gas and $<$N({\it c}-\c3h2)/\EBV$> = 3.3\pm 0.11\times 10^{12}$/mag
averaged over the four sightlines.  However N({\it c}-\c3h2)/N({\it l}-\c3h2) 
$\approx 17$ and $<$N({\it l}-\c3h2))/\EBV$> = 2 \pm 1 \times 10^{11}\pcc$/mag.
This is much smaller than what would be needed for {\it l}-\c3h2\ to carry the
the 5069 \AA\ DIB with which it could be associated on the basis of spectroscopic 
coincidences; \cite{MaiCha+11} quote N({\it l}-\c3h2)/\EBV\ 
$= 5\times10^{14}\pcc$/mag. 

\subsection{Some stuff we ignored or slighted}

The narrow focus of this work prevented discussion of the extensive 
molecular inventory that has recently been discovered and studied in the 
sub-mm regime above 300 GHz with the Herschel satellite under the PRISMAS
Project -- CH\p\ and SH\p; OH\p, O\HH \p\ and OH$_3$\p; OH and 
\HH O; NH, N\HH\ and \ammon; and HF, among others. Some of these, 
for instance OH\p\ \citep{IndNeu+12}, are preferentially found in
regions of very small molecular fraction that may have special relevance
to study of the DIBS.

The spectroscopic summary earlier in this Section can be supplemented by recent 
discussions of interstellar molecules and chemistry, for instance \cite{Tie10}
and \cite{YamWIn11}.  A standard reference to observing techniques in the microwave 
regime is found in \cite{WilRoh+09}.

\section{Systematics}

One general result of molecular spectroscopy is the presence of a small core
group of ubiquitously species having stable (nearly-fixed) abundances 
with respect to each other and \HH.  The group includes OH and \hcop\ 
\citep{LucLis96,LisLuc96,LisPet+10} along with the hydrocarbons CH, 
\cch\ and c-\c3h2\ \citep{LucLis00C2H,GerKaz+11}.  Less-abundant species such 
as {\it l}-\c3h2\ have not been traced in sufficient detail to be included.  
The fixed ratios 
X(CH) = N(CH)/N(\HH)  = $3.5\times 10^{-8}$ and 
X(OH) = N(OH)/N(\HH) $= 1 \times 10^{-7}$ are directly measured at optical/uv 
wavelengths \citep{WesGal+10}.  In turn, OH and \hcop\ have a very nearly fixed ratio 
N(OH)/N(\hcop) = 30-50 at radio wavelengths, providing one of a few direct ties between the 
abundances of species seen in the optical and radio regimes.  Knowing that
X(\hcop) $= 2-3 \times 10^{-9}$ provides a convenient scale with which
to measure the abundances of other species seen in the radio regime.

\begin{table}
  \begin{center}
  \caption{Molecular inventory from  microwave spectroscopy}
  \label{tab1}
 {\scriptsize
  \begin{tabular}{|l|c|c|c|}\hline 
{\bf Family}  & {\bf Detected} & {\bf Search in progress} &  {\bf Upper limit} \\ \hline
 CH & {\it {CH}},\cch, {\underline { {\it c}-C$_3$H}},{\it c}-\c3h2,{\it l}-\c3h2 
  & {\it l}-C$_3$H & 
  \cfh,\cfh$^-$ \\ \hline
 OH & {\it {OH}},\HH O,{\it {CO}},\hcop,\hocp,{\underline {HCO}},\hhco 
    & \HH COH\p & CH$_3$OH \\ \hline
 CN & {\it {CN}},HCN,HNC, \ammon & HC$_3$N,C\HH CN,CH$_3$CN,CH$_3$NC,HNCO 
 & HC$_5$N,N$_2$H\p \\ \hline
 CS & {\it {CS}},SO,\HH S,HCS\p & OCS & SO$_2$,\HH CS \\ \hline
 SiO & SiO & & SiS \\ \hline
 CF\p & {\underline {CF\p}} && \\ \hline
  \end{tabular}
  }
 \end{center}
\vspace{1mm}
 \scriptsize{
 {\it Notes:}\\
  {\underline {species}} = new detection (unpublished); 
  {\it {species}} = also seen  in optical/uv}
\end{table}

Another tie-in between the optical and radio regimes is provided by 
113 GHz measurements of CN, which is always accompanied by HCN and HNC 
in the ratio CN:HCN:HNC = 6.8:1:0.21 \citep{LisLuc01}.  The 
large HCN/HNC ratio is typical of warmer molecular gas (ie 30-50 K) 
as opposed to a dark cloud like TMC-1 at 10 K where all three species 
have comparable abundances (see the table in Figure 5).
Figure 3 uses the CN-family abundances to illustrate the relative stability
of the core group abundances of \hcop\ and \cch\ and the rapid 
increase in the abundances of species outside the core group at 
N(\hcop) $\approx  10^{12}\pcc$, 
corresponding to  N(\HH) $\approx 3\times10^{20}\pcc$.  Shown at left is the 
variation of N(HCN) with respect to N(\hcop) and N(\cch) and at right the
analogous behaviour of CN with respect to \HH.  The abundance of HCN increases
by about a factor 100 over relatively small ranges in N(\hcop) or N(\cch) and
CN is not detected reliably when N(\HH) $< 2-3\times10^{20}\pcc$.

Figure 4 ties the core group and CN-family to the remainder of the observable
molecular inventory using the many absorbing features found along the
 line of sight to B0355+508 (also shown in Figures 2 and 3).  Although this line 
of sight at galactic latitude b = -1.6\degr\ is optically very opaque when 
viewed through the galactic disk, the five most strongly \HH-bearing clouds 
are individually diffuse, all with N(\hcop) $\approx 1.4 \times 10^{12}\pcc$ or 
N(\HH) $\approx 4 \times 10^{20}\pcc$, which is about half that seen toward \bll\ 
where \AV\ = 1 mag.  Although all of the five strong kinematic absorption components 
have about equal molecular content only the two at -11 \kms\ and -17 \kms\ are strong in 
\hhco, CS, \HH S, HCN and \ammon.  The component at -8.5 \kms\ has weak 
absorption in HCN and perhaps in \hhco.  The \ammon\ spectrum at top is complicated 
by hyperfine structure and the component at -4 \kms\ is actually missing.  

Figure 4 shows that species outside the core group appear or disappear together.  
Carbon monoxide is an exception because it is seen almost as ubiquitously as 
the core group species but its abundance is more variable \citep{LisLuc98}.  
Confusingly,
three of the strong absorption features are equally strong in CO emission, 
including those two that show stronger HNC, etc, but CO emission is also
strong in the feature at -8.5 \kms\ having much weaker (but at least some?) HNC.  
The absorption components at -13 \kms\ and -4 \kms\ show obvious CO absorption
and relatively little CO emission; they are examples of the so-called 
``CO-dark molecular gas,'' ie \HH\ that is not well-traced in CO emission.  
However,  emission is absent only in the immediate vicinity of the continuum 
background target.  Both features are strong in emission at other positions
in the nearby sky field \citep{LisPet12}

\section{The molecular inventory from microwave spectroscopy}

Table 1 summarizes the molecular inventory of diffuse clouds as known from  
microwave astronomical spectroscopy.  Also included are species for which there 
are published upper limits, three species whose detection we discussed at 
IAU297 for the first time and species that are currently being sought at the 
VLA.  Species that are also observed in the optical/uv domain are underlined.

As discussed above, a single representative abundance can only be quoted for the few 
species in the small core group of OH, \hcop, CH, \cch\ and {\it c}-\c3h2. Beyond
this, the CN-family, CS, \hhco\ and \ammon\ have been shown to have widely-varying 
abundances and much smaller abundances at smaller \AV, 
N(\hcop) and N(\HH) (Figures 3 and 4).  Yet other species were observed too marginally 
or in too-limited a fashion to make statements about their systematics.  There 
is some evidence that {\it l}-C$_3$H is also in the core group.  

The table in Figure 5 shows comparisons of molecular abundances across various 
chemical families at high and moderate extinction, the former represented by 
molecular emission measurements toward the cyanopolyyne peak in the dark
cloud TMC-1 at \AV\ = 10 
mag \cite{OhiIrv+92,SmiHer+04} and the latter by our measurements toward \bll\ at 
\AV\ = 1 mag.  The abundances toward TMC-1 are subject to various uncertainties
and only differences of a factor two or more between TMC-1 and \bll\ can be
reliably said to be significant.  The fractional abundances of the fiducial species 
toward \bll\ were set by taking their ratios with respect to \hcop, with X(\hcop) 
$= 3\times 10^{-9}$. 

In arranging this table the fiducial species were chosen as those that are the most 
abundant in TMC-1, ie OH, \cch, CN and CS.  Given the differences in physical conditions 
and chemistry in these two regimes it was hardly to be expected that the two sightlines 
would show any similarities at all but in fact the differences are mostly in the details.  
The fiducial species and the core group species have nearly the same relative abundances 
in both the dark and diffuse domains.  We note that:

{\underline{\it Hydroxyl}}. OH and \hcop\ have comparable relative abundances in both regimes.
The diffuse gas has ten times more HCO and sixteen times less \hhco.  The fractional 
abundance of \methanol\ in TMC-1 is well below the reach of the existing upper limit toward
\bll.  The actual degree of molecular complexity in the diffuse gas remains to be seen.

{\underline{\it Hydrocarbons}}.  X(CH)/X(\cch) $\approx 1/3$ toward
TMC-1, making the hydrocarbons somewhat exceptional because the most abundant species
is polyatomic.  Within the range traced it is only marginally clear that the degree 
of molecular complexity is less in the diffuse gas. The abundance of {\it l}-C$_3$H 
should be known soon. A better limit on \cfh\ is needed.

{\underline{\it Cyanogen}}. Toward TMC-1, CN, HCN, HNC and \ammon\ all have comparable 
abundances and that of HC$_5$N is ten times smaller.  In the diffuse gas, CN is about
ten times more abundant than HCN or \ammon\ and X(CN)/X(HC$_5$N) = 1/200. 
The small HNC/HCN ratio toward \bll\ is typical of warmer gas.  The degree of molecular
complexity is definitely lower toward \bll.

{\underline{\it Sulfur}}. The diffuse gas has three times less CS but rather more
SO and \HH S.

\section{What does all this mean for the carriers of the DIBS?}

If nothing else, microwave spectroscopy can directly provide measured interstellar 
abundances for specifically-targeted polar molecules that are proposed as DIB carrier 
candidates, extending far beyond the small set of species that is observable in optical 
spectroscopy.  This was recently done for {\it l}-\c3h2\ and observations will soon 
be taken in a search for C\HH CN, whose cm-wave spectrum is well-known,
in order to limit the abundance of the anion DIB-carrier candidate C\HH CN$^-$ 
\citep{CorSar07} whose microwave spectrum is not yet published.  Other examples will 
certainly follow.

Beyond this, Table 1 notes yet other molecular species, not necessarily putative 
DIB-carriers, that are being specifically targeted in contemporaneous VLA observations 
in order to enhance our reconnaissance of the molecular inventory.  The hope is, that 
with increasing sensitivity and completeness it will be possible 
to draw more general conclusions.  When the overall run of molecular abundances
is known, and the realizable abundances of DIB-carrier candidates are more tightly
constrained, constraints on the line strength are implied, with consequences
for the internal structure of putative carriers.

Wider searches for new molecular species in diffuse molecular gas are
proliferating independent of considerations of the DIBS.  One example is the 
recent 84-115 GHz spectral sweep from which Figure
1 and the three new detections reported here (HCO, {\it c-}C$_3$H and CF\p )
were drawn; upper limits on the abundances of other species whose lower-lying
transitions fall in the 3mm band will follow from this sweep once some remaining 
low-level instrumental artifacts are removed.  Projects such as the PRIMOS 
cm-wave spectral sweep of dense-gas emission toward Sgr B2 at the GBT \citep{ZalSei+13} 
also observe absorption from intervening diffuse clouds, albeit in the central
regions of the Galaxy.

Finally we note that the present chasm between microwave and optical spectroscopy
of diffuse clouds, and any possible doubts about the relevance of microwave 
spectroscopy to the question of the DIBS, will disappear when large optical 
telesopes take spectra of the brighter radio-loud
blazars such as \bll.  Indeed it was hoped to present such spectra
at this meeting, but \bll\ faded below 16th magnitude in the visible before the 
observations could be undertaken during the past observing season.  Our hopes 
for this project, however, remain bright.

\acknowledgements{ This work was funded in part by grant 
ANR-09-BLAN-0231-01 from the French Agence Nationale de la Recherche as part of the SCHISM
project(http://schism.ens.fr/).}



\end{document}